\documentclass[pra,floatfix,twocolumn,superscriptaddress,showpacs,preprintnumbers,longbibliography,nofootinbib]{revtex4-1}

\usepackage[utf8]{inputenc} 
\usepackage{bibunits} % more bibliographies
\usepackage{amssymb,amsfonts,amsmath}
\usepackage{mathtools}
\usepackage{float} 
\usepackage{amsthm}
\usepackage{enumitem}
\usepackage{bbold}
\usepackage[normalem]{ulem}
\usepackage{hyperref} % add hyperlinks to labels
\hypersetup{colorlinks=true, urlcolor=blue, citecolor=blue,linkcolor=blue}
\usepackage[all]{hypcap} % let hyperlinks correctly point to figures rather than their captions; must be loaded after the hyperref package!
\usepackage{multirow}

\usepackage[mathscr]{euscript}
\usepackage{graphicx}% Include figure files
\usepackage{dcolumn}% Align table columns on decimal point
\usepackage{bm}% bold math
\usepackage{psfrag}
\usepackage{epsfig}
\usepackage{multirow}
\usepackage{color}
\usepackage{bbm}
\usepackage[FIGTOPCAP,raggedright,nooneline]{subfigure}
\usepackage{verbatim}
\usepackage{upgreek} % for \tau
\usepackage{physics}
\usepackage{tikz} %circled numbers

\DeclareMathOperator{\diag}{diag}

\newcommand{\T}[1]{\text{#1}}

\newcommand{\ignore}[1]{}

\begin{document}

		\title{Hermitian and Non-Hermitian Topology from Photon-Mediated Interactions}
		
		\author{Federico Roccati}
		\affiliation{Department of Physics and Materials Science, University of Luxembourg, L-1511 Luxembourg}
		
		\author{Miguel Bello}
		\affiliation{Max-Planck-Institut für Quantenoptik, Hans-Kopfermann-Strasse 1, Garching 85748, Germany}
		\affiliation{Munich Center for Quantum Science and Technology, Schellingstraße 4, 80799 München, Germany}
		
		\author{Zongping Gong}
    	\affiliation{Max-Planck-Institut für Quantenoptik, Hans-Kopfermann-Strasse 1, Garching 85748, Germany}
		\affiliation{Munich Center for Quantum Science and Technology, Schellingstraße 4, 80799 München, Germany}
        \affiliation{Theoretical Quantum Physics Laboratory, Cluster for Pioneering Research, RIKEN, Wako-shi, Saitama 351-0198, Japan}

        \author{Masahito Ueda}
        \affiliation{Department of Physics, The University of Tokyo, 7-3-1 Hongo, Bunkyo-ku, Tokyo 113-0033, Japan}
        \affiliation{RIKEN Center for Emergent Matter Science (CEMS), Wako, Saitama 351-0198, Japan}
        \affiliation{Institute for Physics of Intelligence, The University of Tokyo, 7-3-1 Hongo, Bunkyo-ku, Tokyo 113-0033, Japan}

		\author{Francesco Ciccarello}
		\affiliation{Universit{\`a}  degli Studi di Palermo, Dipartimento di Fisica e Chimica -- Emilio Segr{\`e}, via Archirafi 36, I-90123 Palermo, Italy}
		\affiliation{NEST, Istituto Nanoscienze-CNR, Piazza S. Silvestro 12, 56127 Pisa, Italy}

        \author{Aur\'elia Chenu}
		\affiliation{Department of Physics and Materials Science, University of Luxembourg, L-1511 Luxembourg}
	
		\author{Angelo Carollo}
		\affiliation{Universit{\`a}  degli Studi di Palermo, Dipartimento di Fisica e Chimica -- Emilio Segr{\`e}, via Archirafi 36, I-90123 Palermo, Italy}

    \begin{abstract}
       
        Light can mediate effective dipole-dipole interactions between atoms or quantum emitters coupled to a common environment. 
        Exploiting them to tailor a desired effective Hamiltonian can have major applications and advance the search for many-body phases. Quantum technologies are mature enough to engineer large photonic lattices with sophisticated structures coupled to quantum emitters. In this context, a fundamental problem is to find general criteria to tailor a photonic environment that mediates a desired effective Hamiltonian of the atoms.
        Among these criteria, {\it topological} properties are of utmost importance since an effective atomic Hamiltonian endowed with a non-trivial topology can be protected against disorder and imperfections. Here, we find general theorems that govern the  topological properties (if any) of photon-mediated Hamiltonians in terms of both Hermitian and non-Hermitian topological invariants, thus unveiling a \textit{system-bath topological correspondence}. 
        The results depend on the number of emitters relative to the number of resonators. 
        For a photonic lattice where each  mode is coupled to a single quantum emitter, the Altland-Zirnbauer classification of topological insulators allows us to 
        link the topology of the atoms to that of the photonic bath: we unveil the  phenomena of \textit{topological preservation}  and \textit{reversal} to the effect that the atomic topology can be the same or opposite to the photonic one, depending on Hermiticity of the photonic system and on the parity of the spatial dimension. As a consequence,  the bulk-edge correspondence implies the existence of atomic boundary modes with the group velocity opposite to the photonic ones in a 2D Hermitian topological system.
        If there are fewer emitters than photonic modes, the atomic system is less constrained and no general photon-atom topological correspondence can be found. We show this with two counterexamples.
    \end{abstract}

    \maketitle

     The study of topological phases and related phenomena, such as edge states protected against disorder, dates back to the discovery of the quantum Hall effect  in the 80s~\cite{von202040}. 
     Since then, the field of topological phases of matter has expanded  considerably  and today stands out as a leading theme of modern condensed matter physics~\cite{KaneRMP10,qi2011topological} and photonics~\cite{lu2014topological,OzawaRMP2019}.
     This rapid growth was also prompted by the quest for quantum technologies immune to disorder and to the detrimental interaction with an environment~\cite{nayak2008non}. 
     Currently, a new paradigm of topological invariants is under intense investigation within the framework of non-Hermitian physics, a burgeoning research area spanning over  photonics, condensed matter and ultracold atoms~\cite{UedaReview,BergholtzRMP21}.

    Despite considerable efforts in solid-state and photonic settings, the exploration of topological effects in quantum optics -- and especially atom-photon interaction -- is still at an early stage. Some theoretical and experimental works exploited topologically-protected photonic edge modes as channels enabling uni-directional emission~\cite{barik2018topological,mehrabad2020chiral}, excitation/quantum state transfer between quantum emitters~\cite{CiccarelloPRA2011,yao2013topologically,Almeida2016,lemonde2019quantum} and multi-mode entanglement generation~\cite{vega2022topological}.
	Notably, it was predicted in a photonic Su–Schrieffer–Heeger (SSH) model that, being a quantum zero-dimensional defect, an atom itself can seed {\it dressed} bound states that are topologically protected~\cite{BelloSciAdv2019}, the essential properties of which as well as criteria for their occurrence were then derived on a general basis~\cite{LeonfortePRL2021}. Specific investigations of topological dressed states were carried out in other photonic analogues of prototypical topological models such as the  Harper–Hofstadter~\cite{PeterHall21,BelloPRB2023} and the Haldane model~\cite{LeonfortePRL2021,BelloPRB2023} as well as in lossy systems exhibiting non-Hermitian topology~\cite{GongPRL2022,GongPRA2022}. Moreover, atomic emission properties were proposed as sensors of photonic topological phases~\cite{VegaPRA21}. Much of this recent literature is motivated by outstanding technological progress allowing the fabrication of photonic lattices with engineered properties in the form of large periodic  arrangements (1D or 2D) of coupled cavities/resonators  and coherently couple them to a set of controllable atoms/quantum emitters in various experimental platforms such as ultracold atoms~\cite{Krinner2018}, circuit QED~\cite{Liu2017a,SundaresanPRX15,KimPRX2021,ScigliuzzoPRX2022,owens2022chiral} and coupled-resonator optical waveguides~\cite{NoriCROW22}.

    Against the above backdrop, an essential question still remains unanswered: 
    Given a photonic bath with known topological properties, do the atoms coupled to it inherit some  non-trivial topology? If so, how are the  topological symmetry class and phase of the atoms related to those of the field?

    To tackle such a general issue, we adopt  the standard Altland-Zirnbauer classification of topological insulators~\cite{AltlandPRB1997} and consider a general model of photonic lattice weakly coupled to a periodic arrangement of two-level emitters in a way that the total system remains translationally invariant (with a unit cell generally larger than that of the bare photon field). 
    We find general results linking the photonic and atomic topological invariants. We thus unveil a relation between photonic and atomic boundary modes under open boundary conditions, on the basis of the bulk-edge correspondence in the Hermitian case~\cite{ChiuRMP2016}, and also a relation between skin modes or more general bulk anomalous dynamics
    in the non-Hermitian case~\cite{OkumaPRL2020,Lee2019}. 
    The main results of this system-bath topological correspondence are summarized  in Table~\ref{tableResult}.

\begin{table}[t!]
	\centering
	\includegraphics[width=0.95\columnwidth]{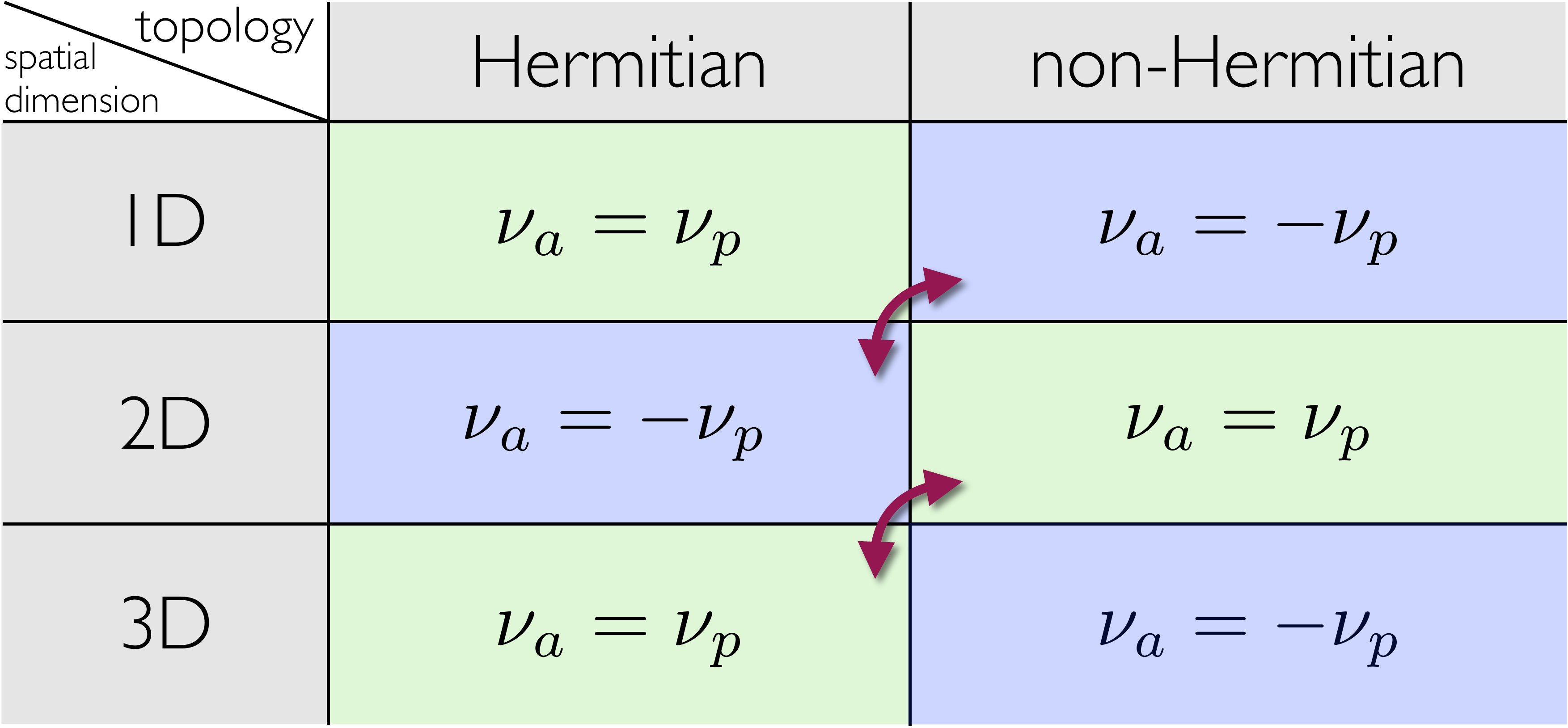}
	\caption{
		{\textit{Main results of the present work.}
			If the degrees of freedom of the system and those of the bath are equal
			the system's topology is preserved ($\nu_a=\nu_p$) or reversed ($\nu_a=-\nu_p$) with respect to that of the bath, according to Eq.~\eqref{niceeq}. Only $\mathbb Z$ phases are included in the table, as in the $\mathbb Z_2$ case topology is always preserved.  The red arrows connect those cases that are linked by the Hermitian-non-Hermitian correspondence~\cite{Lee2019}. The resonance condition $\omega_e=\omega_0$ is assumed (c.f.~main text).
		}
	} 
	\label{tableResult}
\end{table}

	\begin{figure}[t!]
			\centering
			\includegraphics[width=0.95\columnwidth]{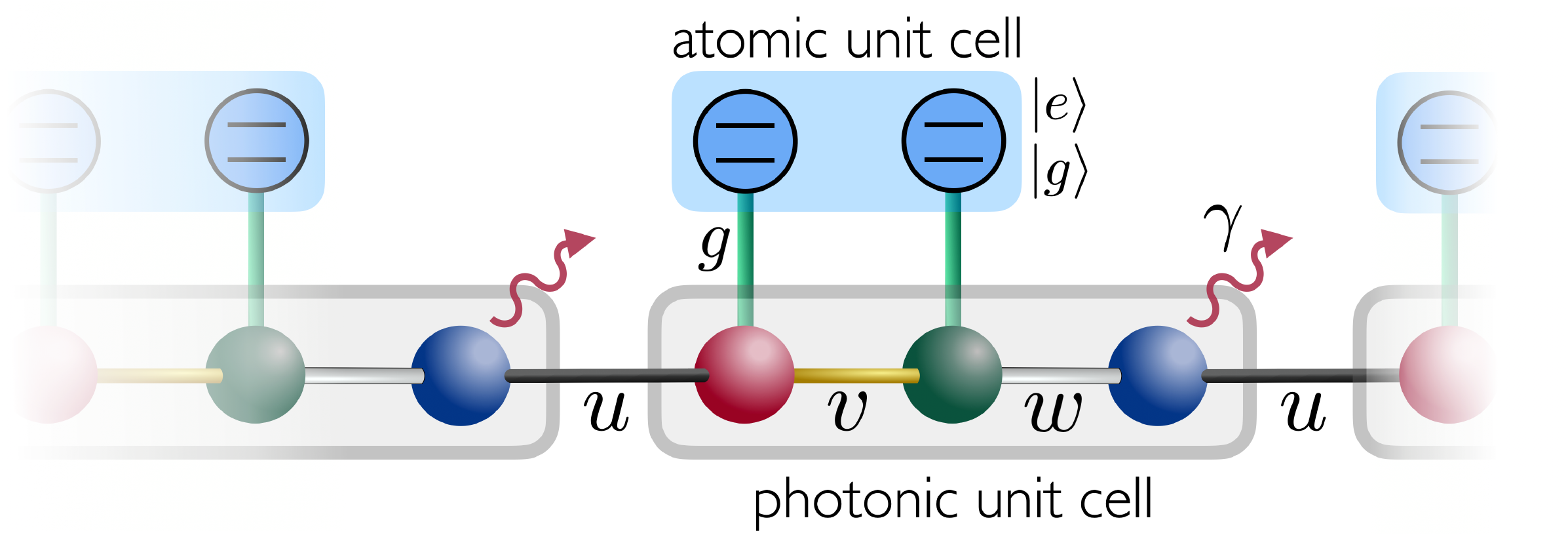}
			\caption{
				{\textit{Setup.}
				Two-level ($\ket{g}$, $\ket{e}$) quantum emitters (QEs) coupled to a photonic lattice.
                The full light-atom system is translationally invariant.  
                The figure  shows a specific 1D example with two emitters in a photonic unit cell with $N_b=3$ resonators (red, green and blue), for clarity. 
                A more general situation is discussed in the main text.
                The intracell couplings are $v$ and $w$, the intercell one is $u$, and the third resonator (blue) is lossy with rate $\gamma$. 
				Translational invariance of the light-atom Hamiltonian is achieved, e.g., by coupling two QEs to the first two resonators of each photonic unit cell. This way the atomic Hamiltonian features a unit cell of dimension 2 (cyan). 
                This scheme is naturally  applicable to superconducting cavity systems where the transmission line plays the role of the photonic bath and superconducting qubits play the role of two-level atoms.
				}
			} 
				\label{fig0}
			\end{figure}

	\section*{Results}

        \subsection*{System-Bath Topological Correspondence}

        Before diving into the specific quantum optical setup under our consideration, we present our result in a general setting. 
        
        Consider a multipartite quantum system with free Hamiltonian $\hat H_e = \omega_e\sum_i \hat s_i^\dagger \hat s_i$, with $\hat s_i^{(\dagger)}$ being the annihilation (creation) operator of the $i$th system's excitation. 
        Suppose this system is coupled to a bath. We assume that the bath Hamiltonian is number conserving, quadratic and translationally invariant. Then, under periodic boundary conditions, it can be written in the second quantized form as $\hat H_p=\sum_{\textbf{k}}
		      \hat B_\textbf{k}^\dagger\,
		      H_p(\textbf{k})
		      \hat B_\textbf{k}$. Here $\textbf{k}$ is the quasi-momentum, which is a good quantum number because of translational invariance. The matrix $H_p(\textbf{k})$ is the single particle Hamiltonian in Fourier space, whose eigenvalues yield the bath's band structure, centered around the energy $\omega_0$. Finally, $\hat B_\textbf{k}=(\hat b_{\textbf{k},1},\hat b_{\textbf{k},2},\ldots)^{\text{\sc T}}$ where $\hat b_{\textbf{k},j}$'s are the Fourier transform of the bath annihilation operators in real space $\hat b_{\textbf{r},j}$, i.e., $\hat b_{\textbf{k},j}\propto\sum_{\textbf{r}} e^{i \textbf{k}\cdot \textbf{r}}\hat b_{\textbf{r},j}$.        
        We consider a weak system-bath interaction  given by a coherent exchange of excitations, that is $\hat H_\text{int}= g \sum_i(\hat s_i\hat b_i^\dagger+\text{H.c.})$.

        Assuming 
        the dynamics is Markovian, the degrees of freedom of the bath can be traced out. When the system bare frequency $\omega _e$ does not belong to the bath spectrum, this procedure leads to an effective \textit{dressed} system Hamiltonian $\hat H_a\sim \hat H_e + %\hat 
        g^2\hat G_p(\omega_e)$, where the latter is the resolvent operator of the bath~\cite{Economou2006}.  

        Our results relate the topological properties of the bare bath and  of the system dressed by the bath, establishing  a \textit{system-bath topological correspondence}. 
        By setting $\omega_e=\omega_0$, both Hamiltonians enjoy the same symmetries. In addition, when the number of degrees of freedom of the system and those of the bath are equal to each other, we unveil the phenomena of\textit{ topological preservation} and \textit{reversal} on the basis of the Altland and Zirnbauer classification~\cite{AltlandPRB1997}. Namely, the topological invariants $\nu_{a(p)}$   of $\hat H_{a(p)}$ are related as
        \begin{equation}\label{niceeq}
            \nu_a=
            \begin{cases}
            \nu_p & \text{for $\mathbb Z_2$ phases,}\\
            \nu_p(-1)^{D+\mathfrak{h}} & \text{for $\mathbb Z$ phases.}
            \end{cases}
        \end{equation}
        Here $D$ is the spatial dimension while $\mathfrak{h}$ is 1 (0) if $\hat H_p$ is (non-)Hermitian. 
        Integer topological invariants of the bath are preserved or reversed by the system, according to the dimension and Hermiticity.
        $\mathbb Z_2$ protected phases of the bath are instead always inherited by the system.

        This result is in accordance with the Hermitian-non-Hermitian correspondence introduced in Ref.~\cite{Lee2019}. Therefore, the boundary mode of a $D+1$ 
        Hermitian topological system characterized by the topological invariant $\nu$ can be mapped into a $D$ 
        non-Hermitian system on a closed manifold, with identical topological invariant. Indeed, we show how topological preservation (reversal), i.e.,~$\nu_a=\nu_p$ ($\nu_a=-\nu_p$), occurs in  $D+1$ dimensional 
        Hermitian systems as well as 
         non-Hermitian ones in $D$ dimensions, see Table~\ref{tableResult}.
        
	   \subsection*{ Quantum Optical System and Hamiltonian
    }

      In the commonly-studied weak-coupling and Markovian regime, photons mediate second-order interactions between the atoms which can be described by an effective many-body atomic Hamiltonian~\cite{BelloSciAdv2019}. We study the topological properties of the latter Hamiltonian, showing that these depend in particular on the detuning between the atomic frequency and the mean photonic frequency (the latter is typically located at the middle of the central photonic bandgap).
		
	   The setup we consider,  a specific instance of the general aforementioned setting, is illustrated in Fig.~\ref{fig0}. It comprises $N_e$ two-level quantum emitters (QEs), each with the ground state and an excited state, $\ket{g}$ and $\ket{e}$, separated by the Bohr frequency $\omega_e$. 
	   The QEs are locally coupled to a translationally invariant photonic lattice implemented by  coupled single-mode resonators. The unit cell of the lattice contains $N_b$ resonators. 
	   Hence, there are (generally) as many sublattices as  photonic bands.
        
        The system is modeled by the Hamiltonian
    	$	\hat H=\hat H_e+\hat H_p+\hat H_\T{int}\,.$
	    The free atomic Hamiltonian reads $\hat H_e = \omega_e\sum_n\hat\sigma_n^\dagger \hat\sigma_n$, 
        where $\hat \sigma_n=\ket{g}\!_n\!\!\bra{e}$. 
	    Under periodic boundary conditions (BCs), the bare photonic Hamiltonian can be expressed in terms  of bath modes with a definite quasi-momentum $\textbf{k}$ as
	       \begin{equation}\label{HB}
		      \hat H_p= \sum_{\textbf{k}\in \T{BZ}}
		      \hat A_\textbf{k}^\dagger\,
		      H_p(\textbf{k})
		      \hat A_\textbf{k}\,,
	       \end{equation}
	    where ``BZ" stands for the first Brillouin zone, 
		$\hat A_\textbf{k}^\dagger=(\hat a_{\textbf{k},1}^\dagger,...,\hat a_{\textbf{k},N_b}^\dagger)$, $\hat{a}_{\textbf{k},s}$'s are the bosonic annihilation operators of the field's normal modes, and $H_p(\textbf{k})$ is the $N_b\times N_b$ Bloch Hamiltonian matrix (see Methods). We label the bare resonator frequency by $\omega_0$ and set $\omega_0=0$ as the reference energy.
	    The last term in the total Hamiltonian describes the interaction between QEs and the field according to the usual rotating-wave approximation,
	       \begin{equation}\label{interaction}
		      \hat H_\T{int} = g \sum_{n=1}^N 
            \sum_{s\in{\mathcal C}} 
                \hat \sigma_{ns}^\dagger \hat a_{ns} + \T{ H.c} \, ,
	       \end{equation}
        where $\mathcal C\subseteq\{1,...,N_b\}$ is the set of sublattices coupled to QEs.
    Here, $\hat a_{ns}$ is the real-space annihilation operator of the resonator located in the $n$th unit cell, belonging to the $s$-sublattice ($s\in\{1,...,N_b\}$), and $N$ is the total number of unit cells.  
	   The atomic operator in Eq.~\eqref{interaction} has two indices in order to specify the resonator it couples to.

	   In the following, we will consider a translationally invariant arrangement of emitters, with  the periodicity equal to or larger than that of the photonic lattice.
	   Provided that $\omega_e$ is in a photonic bandgap and the coupling constant $g$ is small (small compared to the spectral distance between $\omega_e$ and the bath's bands), the bath will induce effective coherent interactions between the emitters~\cite{TudelaNatPhot2015,DouglasNatPhot2015,BelloSciAdv2019}.
	   In this weak-coupling regime, the atomic and photonic degrees of freedom do not mix much, and it is meaningful to analyze the topological properties of the atomic subsystem alone.
		
		\subsection*{One-emitter-per-resonator case:  topological preservation and reversal}

        In the one-emitter-per-resonator case, 
        the Bloch Hamiltonian of the entire system is
        \begin{equation}
            H(\mathbf k) = \begin{bmatrix}
            \omega_e I & g I \\ g I & H_p(\mathbf k)
            \end{bmatrix} \,,
            \label{Hes}
        \end{equation}
        where $I$ is the $N_e$-dimensional identity matrix.
        Remarkably, the entire atom-light Hamiltonian is  topologically trivial (see Methods).
        By means of standard perturbation theory to second order, the effective atomic Hamiltonian  is obtained (see Methods), and depends on the resolvent operator of the bath as
		\begin{equation}\label{AtomicBlochPiIdentity}
			H_a(\mathbf{k}) = \omega_e + \frac{g^2}{\omega_e - H_p(\mathbf{k})} \,.
    	\end{equation} 
    	The real space form, $\hat H_a$, is recovered by means of the inverse Fourier transform.
    
    	We find that, depending on the spatial dimension, the symmetries, and whether the photonic Hamiltonian is Hermitian or not, the topology of this effective system may be either preserved or reversed, at least for the fundamental symmetry classes. 
        We provide here the general results in the Hermitian and non-Hermitian cases as  summarized in Table~\ref{tableResult}, together with prototypical examples.

		\subsubsection*{Hermitian case}

            The identification of the 10 fundamental symmetry classes was first made by  Altland and Zirnbauer (AZ)~\cite{AltlandPRB1997}.
            On this basis, the topological classification of Hermitian non-interacting systems (insulators and superconductors) has been later carried out~\cite{SchnyderPRB2008,Kitaev2009}.
        	Only five of these symmetry classes are relevant for the number conserving 
         Hamiltonians we consider~\cite{OzawaRMP2019}, see Table~\ref{table}.
         The classification is based on time-reversal symmetry (TRS or $T$), particle-hole symmetry (PHS or $C$), and chiral symmetry ($S$)~\cite{AltlandPRB1997,ChiuRMP2016,OzawaRMP2019}. 
         Their explicit action on the system Bloch Hamiltonian $H(\textbf{k})$ is given by
		\begin{align}
		    T H(\textbf{k}) T^{-1} 
		    &= H(-\textbf{k}),
      \label{TRS}
		    \\
		    C H(\textbf{k}) C^{-1}
		    & = - H(-\textbf{k}),
      \label{PHS}
		    \\
		    S H(\textbf{k}) S^{-1}
		    & = - H(\textbf{k}).
      \label{CS}
		\end{align}
		Both $T$ and $C$ are antiunitary operators, i.e., $T=U_{\text{\tiny TRS}}K$ and $C=U_{\text{\tiny PHS}}K$, where $U_{\text{\tiny TRS}}$ and $U_{\text{\tiny PHS}}$ are unitaries and $K$ denotes complex conjugation. 
        Applying these to $H_a(\mathbf{k})$, one can show  that TRS is never broken,  
		\begin{equation}\label{trs}
			H_a(\textbf{k}) \T{  has  TRS}
			\,\Leftrightarrow\,
			H_p(\textbf{k}) \T{  has  TRS}
		\end{equation}
		for any atomic frequency in a bandgap. In turn, PHS, and therefore chiral symmetry~\cite{ChiuRMP2016}, can be broken. Indeed
		\begin{equation}\label{phs}
			H_a(\textbf{k}) \T{  has  PHS}
			\,\Leftrightarrow\,
			H_p(\textbf{k}) \T{  has  PHS}
		\end{equation}
		\textit{only if} $\omega_e = 0$, the same holding for chiral symmetry. Conversely, in absence of photonic symmetries, no new symmetry can be generated at the atomic level. 
		
        Equations~\eqref{trs} and~\eqref{phs} show that on resonance ($\omega_e=0$), $\hat H_a$ and $\hat H_p$ belong to the same symmetry class; off resonance, the following transitions of symmetry classes take place when going from $\hat H_p$ to $\hat H_a$: AIII $\rightarrow$ A, BDI $\rightarrow$ AI and D $\rightarrow$ A, see Table~\ref{table}.
        \begin{table}[t!]
			\centering
			\includegraphics[width=0.9\columnwidth]{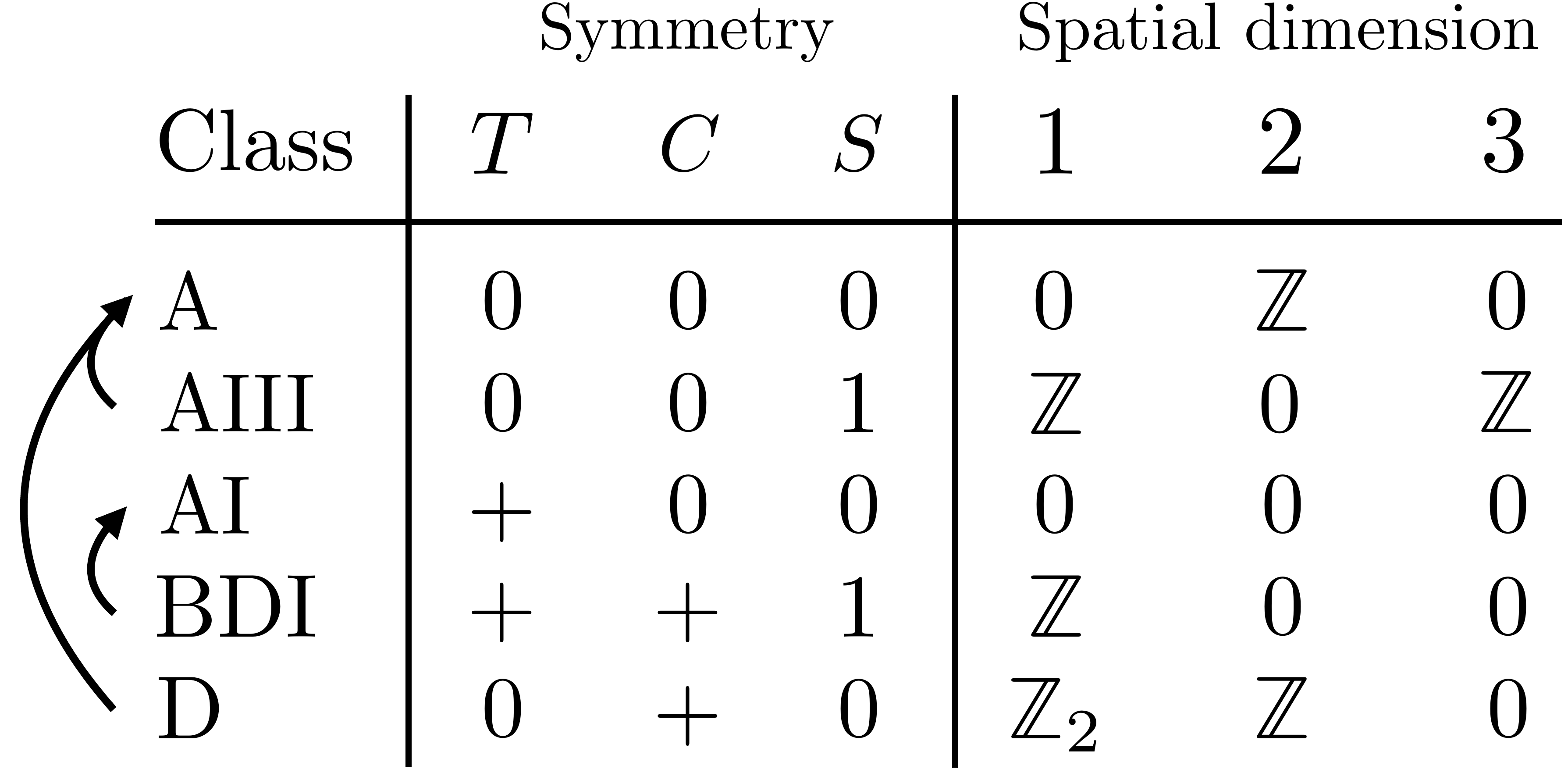}
			\caption{
				{\textit{Periodic table of topological insulators.}
				Altland-Zirnbauer classification of Hermitian photonic topological insulators based on time reversal ($T$), particle-hole ($C$) and chiral ($S$) symmetries, whose presence (absence) is highlighted by $1$ or $+$ ($0$). The $+$ sign indicates that $T$ (or $C$) square to 1, $T^2=1$~\cite{AltlandPRB1997}.
                The black arrows highlight the  off-resonant transitions between photonic and atomic symmetry classes.
                The entries $\mathbb{Z}$, $\mathbb{Z}_2$, $0$ under the spatial dimension represent the possible value of the appropriate topological invariant. 
				The table is adapted from Ref.~\cite{ChiuRMP2016}.
				}
			} 
				\label{table}
			\end{table}
		Within the same symmetry class, topologically distinct  phases are characterized by  different values of the topological invariant (e.g., Zak phase, Chern number, Chern-Simons invariants), which we generally label as $\nu_l$, with $l=p,a$ referring to photonic and atomic subsystems, respectively. 
		According to the bulk-edge correspondence, $\nu_l$ gives the number of edge modes in the system with open BCs, the trivial phase being the one with a zero topological invariant~\cite{ChiuRMP2016}. 

        To ensure that the PHS and chiral symmetry are inherited, we focus on the case  $\omega_e=0$, so that $H_a(\textbf{k})=-g^2H_p(\textbf{k})^{-1}$ with both $H_a(\textbf{k})$ and $H_p(\textbf{k})$ gapped across $0$.

        In the one-emitter-per-resonator case, our main result is that \textit{Hermitian topology is preserved ($\nu_a=\nu_p$) for $\mathbb{Z}_2$ phases in any spatial dimension and for $\mathbb{Z}$ phases in odd dimensions} (in particular $1$D and $3$D). 
        \textit{A topological reversal ($\nu_a=-\nu_p$) occurs for $\mathbb{Z}$ phases in even dimensions} (see Methods for the proof). 
        The topological reversal in particular has direct observable consequences on the basis of the bulk-boundary correspondence (as we discuss later on).

		\begin{figure}[t!]
			\centering
			\includegraphics[width=\columnwidth]{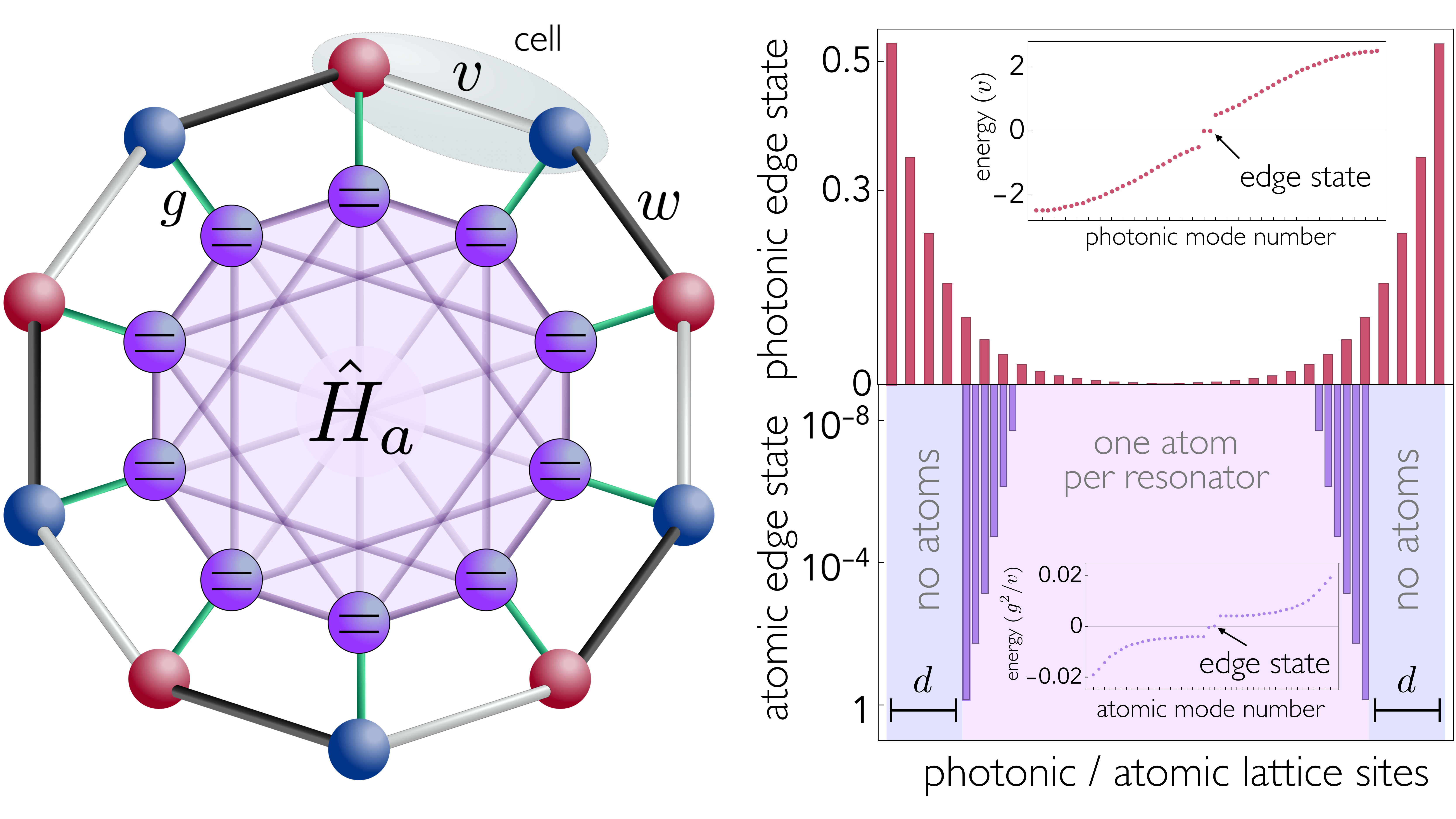}
			\caption{\textit{Hermitian topological preservation.} 
			Left: photonic SSH lattice with staggered couplings $v$ and $w$. Each QE is coupled to each resonator with coupling strength $g$. The form of the mediated QEs' Hamiltonian $\hat H_a$ is sketched in purple. Open BCs for the atomic system  are obtained by removing  QEs, leaving the photonic lattice translationally invariant. 
			Right: Absolute value of the wave function of the photonic edge states with $N=60$  resonators (top), and 
            atomic edge states with $N_e = 44$ emitters (bottom) coupled to a periodic SSH lattice with $N$ resonators (top), $d=8$ sites.
            The atomic edge states have a non-zero weight mostly on the first and the last site (notice the logarithmic scale). The non-zero amplitude on the rest of the sites is a finite-size effect.
			The insets show the photonic (atomic) energy spectra under open BCs in units of $v$ ($g^2/v$). 
					Parameter values: $w=1.5v$, $g=0.1 v$, $\omega_e=0$.
            }
			\label{fig1} 
		\end{figure}

		\textit{Topological preservation---} As a minimal example of Hermitian topological preservation, we consider  the case of  QEs coupled to an SSH lattice, see Fig.~\ref{fig1}, whose Hamiltonian reads
        \begin{equation}
            \hat H_p = \sum_n v\, \hat a_{n,1}^\dagger\hat a_{n,2} + w\,\hat a_{n,2}^\dagger\hat a_{n+1,1} +{\rm H.c.}
        \end{equation}
        This model belongs to the BDI class, hosting $\mathbb Z$ phases~\cite{BelloSciAdv2019}.
		On resonance, the effective Hamiltonian between QEs preserves chiral symmetry and therefore, according to our result, possesses the same topology as the underlying photonic lattice.
        This can be verified through the bulk-edge correspondence. Indeed, by introducing  atomic boundary conditions, which is achievably considering a finite array of QEs in a larger periodic photonic SSH lattice,
        the atomic part supports topological edge states in the non-trivial phase---despite the high connectivity of the mediated interactions.

          \begin{figure*}[ht!]
            \centering
            \includegraphics{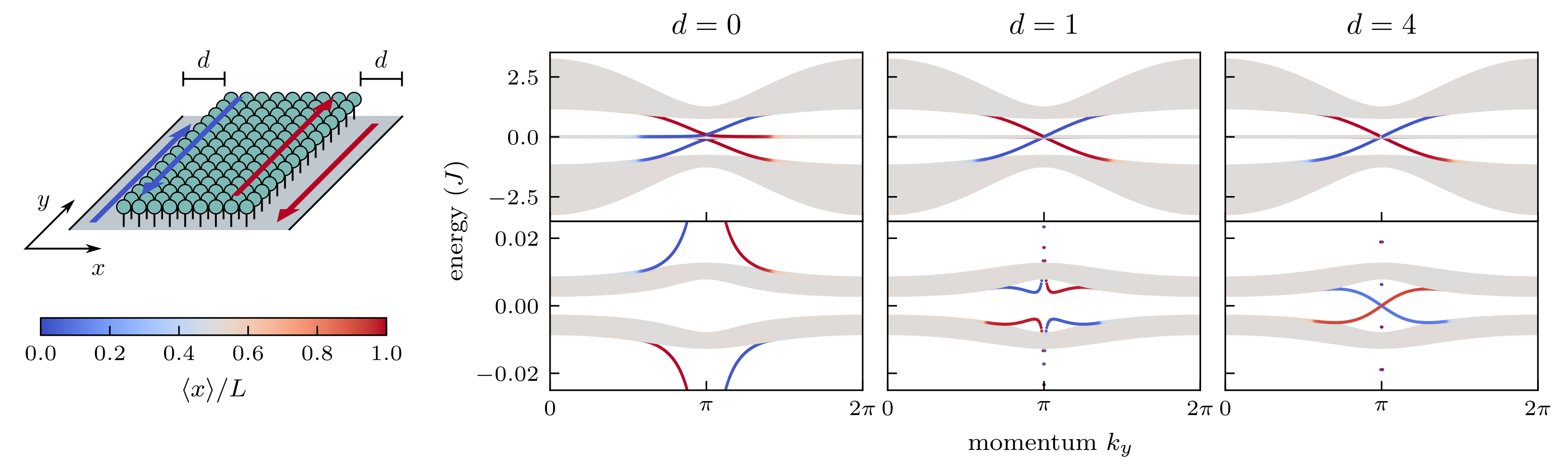}
            \caption{\textit{Hermitian topological reversal.} 
            Upper left: sketch of a setup showing  Hermitian topological reversal. QEs (green spheres) coupled to a QWZ 2D photonic lattice (gray plane), Eq.~\eqref{QWZbloch}. Photonic open boundary conditions  are imposed only along the $x$ direction. The parameter $d$ measures the thickness (in photonic unit cells) of the outer stripes, separating the photonic and atomic boundaries. It 
            serves to interpolate between the case of a system full of emitters ($d=0$), and that of a finite array of QEs in a periodic array of resonators along the $x$ direction at fixed $y$ ($d\gg0$).
            Right:
            energy spectrum of the full system made of quantum emitters coupled to a QWZ bath with open (periodic) boundary conditions along the $x$ ($y$) direction.
            Each emitter is coupled  to one bath site, except for two stripes of size $d$ (unit cells) along the edges of the bath. 
            The plots in the top row highlight the photonic spectrum, the atomic one being very close to zero (zoom in the bottom row; note the different energy scales).
            The colors denote the degree of localization along the  $x$ direction according to the legend (bottom left). 
            States corresponding to grey energies are delocalized, while those corresponding to red (blue) energies are localized at the right (left) boundary.
            The triviality of the full light-matter system is made patent by the absence of edge states in the gap around the resonant frequency (leftmost plot, $d = 0$).
            The effects of the topological reversal are apparent for large $d$ (rightmost plot): focusing on the boundary modes in, e.g., the right boundary (red), the photonic and atomic modes have opposite group velocities.   
            The same holds true for the blue mode.
            Parameter values: $L = 50$ unit cells along the $x$ direction, $u=1.2$, $\omega_e = 0$ and $g=0.1J$.
            The isolated points that can be appreciated in the lower figures for the cases $d = 1$ and $d = 4$ are due to the discretization of the (in fact, continuous) momentum along the $y$ direction chosen for the numerical computations. They correspond to the photonic edge states.
            }
            \label{fig2} 
        \end{figure*}

        \textit{Topological reversal---} The Hermitian topological reversal occurs in two dimensions. To show its implications, we consider QEs coupled to a 2D Chern insulator described by the Qi-Wu-Zhang's (QWZ) model~\cite{QiPRB2006} (class A). The Bloch Hamiltonian $H_p (\mathbf k)$ for this model reads
        \begin{multline}\label{QWZbloch}
		    H_{\text{\tiny QWZ} }(\mathbf k) = 
            J\sin(k_x) \sigma_x + J\sin(k_y) \sigma_y \\ + J[u+ \cos(k_x) + \cos(k_y)] \sigma_z \,,
		\end{multline}
        where $\sigma_\alpha$ ($\alpha \in \{x,y,z\}$)  are the Pauli matrices. Assuming open boundary conditions  along the $x$ direction only, this system supports chiral photonic boundary modes propagating in opposite ways along $y$, whenever $|u|<2$.
        The topological reversal (see Table~\ref{tableResult}) predicts opposite photonic and atomic Chern numbers, resulting in atomic boundary modes featuring excitons which travel with opposite velocities with respect to their photonic counterparts. This is indeed the case when considering a finite array of QEs along the $x$ direction, c.f.~Fig.~\ref{fig2}.  When the set of QEs lies deep in the photonic bulk,  each photonic boundary mode has a corresponding  atomic one but  with opposite chirality on the same boundary.

    \subsubsection*{Non-Hermitian case}

        When the Hamiltonians are allowed to be non-Hermitian, the number of fundamental symmetry classes increases from 10 (AZ classes) to 38 (Bernard-LeClair classes)~\cite{Bernard2002,KawabataPRX2019}. Here we focus only on  a subclass of the latter that naturally generalizes  the former. A crucial difference is that  complex conjugation  is no longer equivalent to  transposition. The appropriate 
        non-Hermitian equivalent of Eqs.~\eqref{TRS}-\eqref{CS} gives the symmetry constraints of the non-Hermitian AZ classes as~\cite{KawabataPRX2019}
        \begin{align}
		    U_\text{\tiny TRS} H^*(\textbf{k}) U_\text{\tiny TRS}^{-1}
		    &= H(-\textbf{k}),
		    \\
		    U_\text{\tiny PHS}  H^{\text{\tiny T}}(\textbf{k}) U_\text{\tiny PHS}^{-1}
		    & = - H(-\textbf{k}),
		    \\
		    S H^\dag(\textbf{k}) S^{-1}
		    & = - H(\textbf{k}).
        \label{NHCS}
		\end{align}
        There is also a dual of these classes, called non-Hermitian AZ$^\dag$~\cite{KawabataPRX2019}, for which the symmetry constraints read
        \begin{align}
		    U_\text{\tiny TRS} H^{\text{\tiny T}}(\textbf{k}) U_\text{\tiny TRS}^{-1}
		    &= H(-\textbf{k}),
		    \\
		    U_\text{\tiny PHS}  H^{*}(\textbf{k}) U_\text{\tiny PHS}^{-1}
		    & = - H(-\textbf{k}),
		    \\
		    S H^\dag(\textbf{k}) S^{-1}
		    & = - H(\textbf{k}).
		\end{align}
        As highlighted in Ref.~\cite{Lee2019}, the topological classification of a Hermitian AZ class in $D$ dimensions coincides with that of the corresponding 
        non-Hermitian AZ (AZ$^\dag$) class in $D+1$ ($D-1$) dimensions.
        The topological preservation and reversal we unveil here are fully consistent with this  Hermitian-non-Hermitian correspondence (see Methods).

        In the one-emitter-per-resonator-case, our main result for the  non-Hermitian AZ (AZ$^\dagger$) classes is that \textit{non-Hermitian topology is always preserved except for $\mathbb Z$ phases in odd dimensions} (see Methods for the proof).
        We discuss here two setups illustrating the non-Hermitian topological reversal and preservation, respectively.

        \textit{Topological reversal---}  This is  %a 
        especially important  for 1D systems, such as those considered in Refs.~\cite{RoccatiOptica2022, GongPRL2022}.
		In 1D, a point gap spectrum characterized by a nontrivial winding number $\nu$ is the topological origin of the non-Hermitian skin effect~\cite{OkumaPRL2020}.
		The sign of $\nu$ can in general be related to the boundary on which the skin states accumulate. Note that  there are also more sophisticated symmetry-protected versions, such as the $\mathbb{Z}_2$ skin effect protected by TRS$^\dag$ discussed in Ref.~\cite{OkumaPRL2020}. This provides an example where the spectral winding always vanishes but with a nontrivial $\mathbb{Z}_2$ number that signals a pair of skin modes localized at both boundaries.
        Nevertheless,
		a non-zero winding number is typical of non-reciprocal models and generally emerges as a combination of broken TRS and dissipation~\cite{ClerkSciPostPhysLectNotes2022}.
  
		The topological reversal entails that photonic skin states on one edge correspond to atomic skin states on the opposite edge. 
		Thus, by locally coupling QEs to all sites of a non-reciprocal lattice, the QEs themselves are a \textit{reversed} non-reciprocal system.

\begin{figure}[t!]
			\centering
			\includegraphics[width=0.95\columnwidth]{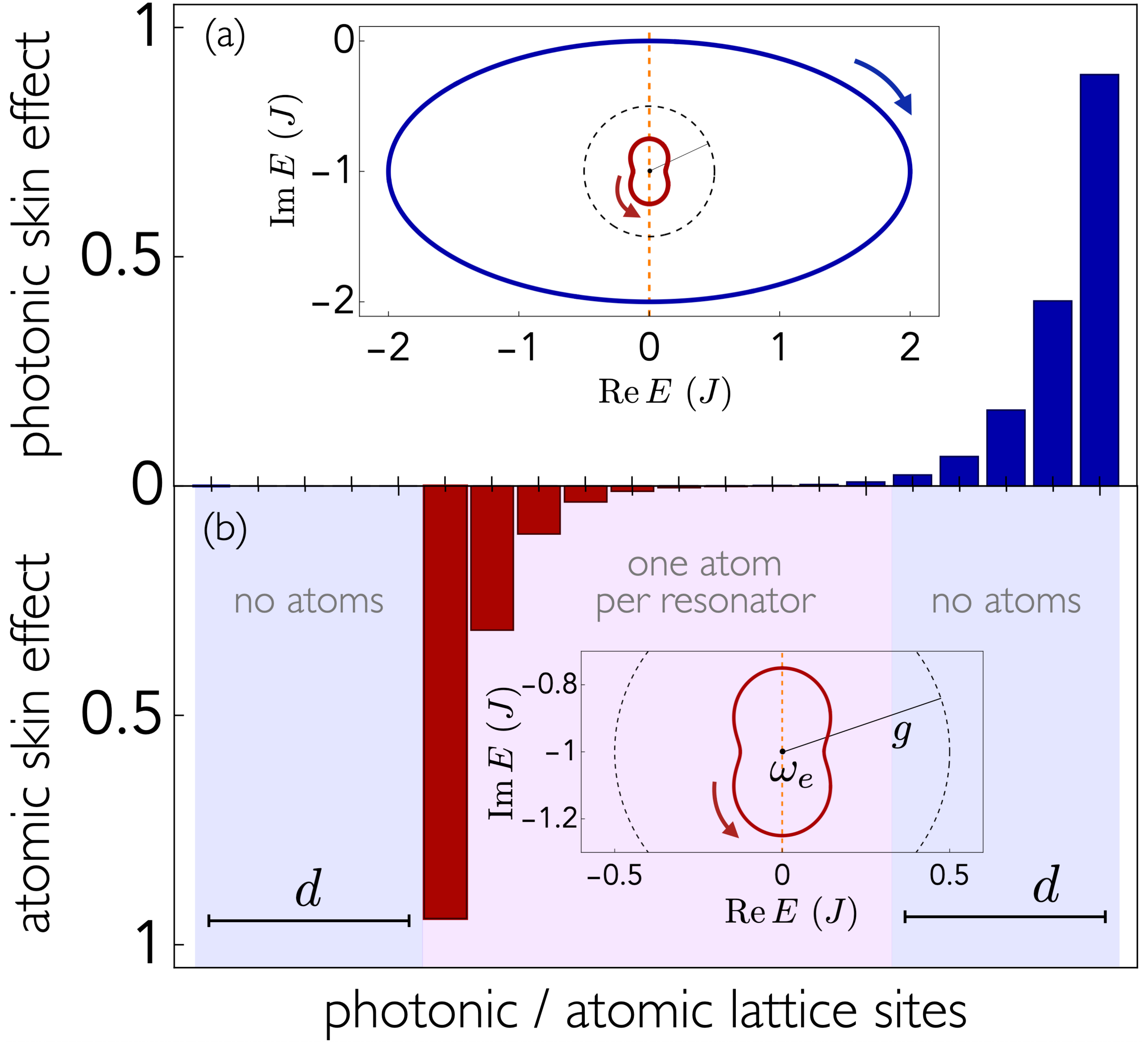}
			\caption{
				{\textit{Non-Hermitian topological reversal.} (a) In a bare photonic Hatano-Nelson model,  Eq.~\eqref{HN} with $N=20$ resonators, photonic skin states accumulate on the right edge. (b) Coupling   emitters to the same photonic lattice,  under periodic BCs, the atomic skin states accumulate on the left edge. Both figures display the normalized average of all skin modes $\ket{\psi_i}$,  $\sum_i|\!\braket{n}{\psi_i}\!|^2$, with $\ket{n}$  being the lattice site.  Atomic open BCs, inducing the skin effect, are taken by removing $2d=10$ QEs (outer violet stripes) while keeping the photonic structure periodic. 
                The top inset shows the  photonic (blue) and atomic (red) complex spectrum under periodic BCs in units of $J$. The bottom inset is a zoom of the top one. We find opposite windings, witnessing the topological reversal. The vertical dashed orange axis is $\Re E = \Re \omega_e$. The dashed black circle with center $\omega_e$ illustrates  the strength of the atom-photon interaction $g$ (the radius). This is highlighted because the reversal in 1D can be described by a circular inversion with respect to this circle~\cite{schwerdtfeger2020geometry}. 
                Parameters: $\delta=0.5 $, $g=0.5J$, $\omega_e=- iJ$.
                }
			}
			\label{fig3} 
		\end{figure}

		As in Ref.~\cite{GongPRA2022}, 
        we consider a photonic Hatano-Nelson 1D array with non-reciprocal right $J_R = J(1 + \delta)$ and left $J_L = J(1 - \delta)$ couplings, and uniform local dissipation $\gamma = 2 \delta J$,
		\begin{equation}\label{HN}
		\hat H_p = \sum_n J_R \hat a_{n+1}^\dagger\hat a_{n} + J_L \hat a_{n}^\dagger\hat a_{n+1} -i\gamma a_{n}^\dagger\hat a_{n}\,
		\end{equation}
		with  QEs coupled to all modes. 
        Since there is only one resonator per unit cell, we dropped the sublattice index.
        Under open BCs, the photonic skin modes accumulate onto the right for $\delta>0$. The atomic periodic system possesses reversed topology (opposite windings), and therefore its skin modes accumulate to the left (see Fig.~\ref{fig3}).

\begin{figure}[t!]
			\centering
			\includegraphics[width=0.95\columnwidth]{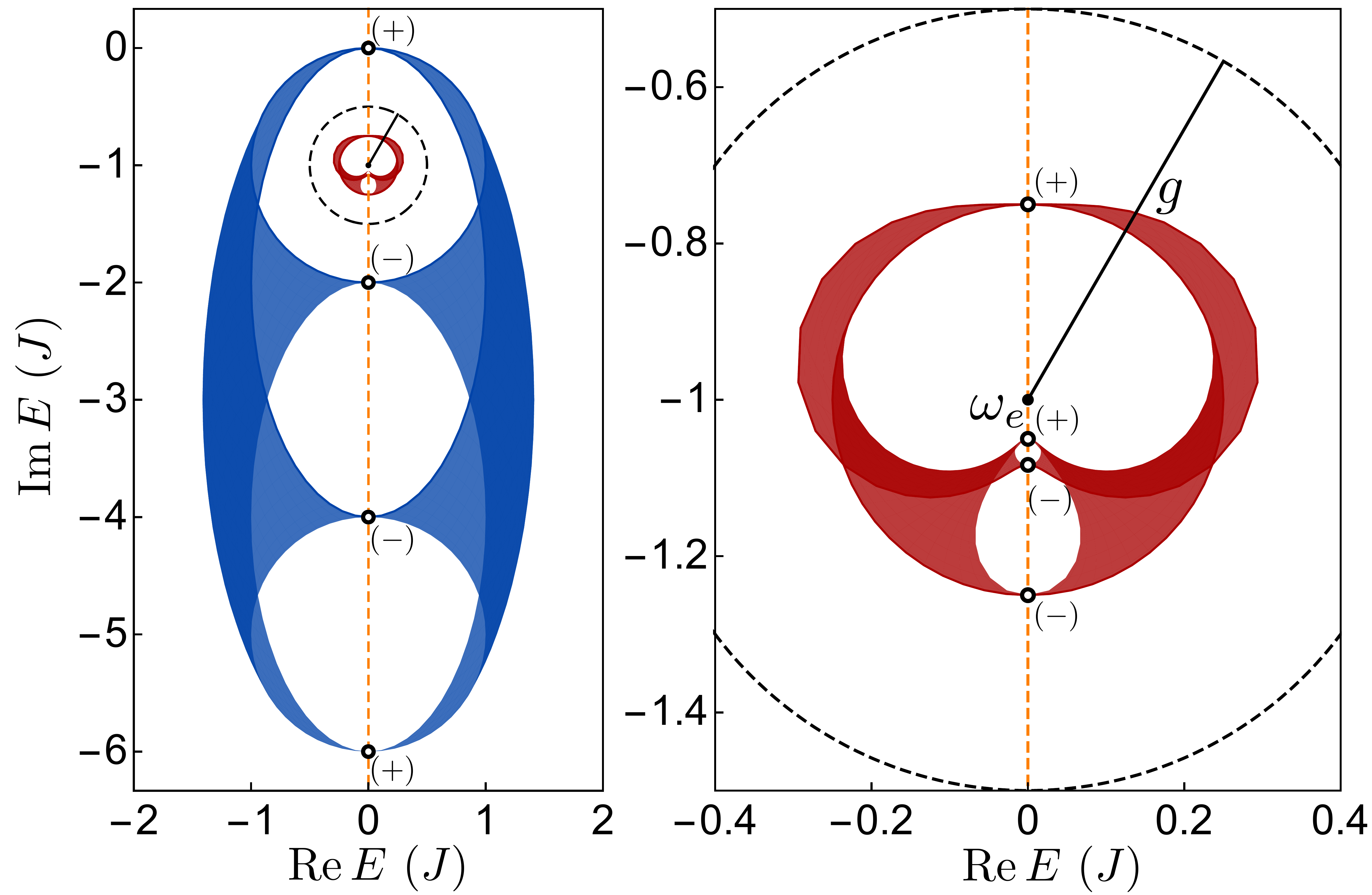}
			\caption{
				{
                \textit{Non-Hermitian topological preservation.} Spectrum of the 2D chiral symmetric non-Hermitian photonic lattice (\ref{2DAIII}) (blue) and that of the coupled emitters spectrum (red) obtained from Eq.~(\ref{AtomicBlochPiIdentity}). The right panel is the zoom-in of the left one. The dots illustrate the Dirac cones with either $(+)$ or $(-)$ chiral charge.  According to Eq.~(\ref{AtomicBlochPiIdentity}), each photonic Dirac cone above (below) $\omega_e$ is mapped to an atomic Dirac cone above (below) $\omega_e$ with the same chiral charge. The topological preservation follows from the fact that the total chiral charge above $\omega_e$ gives the topological number. Dark-shaded areas are swiped twice as $\textbf{k}$ varies in the BZ. Here $\omega_e=-iJ$ and $g=0.5J$.
                }
             }
			\label{fig:NH2D} 
		\end{figure}

        \textit{Topological preservation---} Topology may also be preserved in the non-Hermitian case. As a simple example, we consider a chiral symmetric non-Hermitian photonic bath in two dimensions. From the perspective of Hermitian-non-Hermitian correspondence~\cite{Lee2019}, such a system mimics the gapless surface states of three-dimensional chiral topological insulators~\cite{Hosur2010}. Accordingly, it can be characterized by the net chiral charge of Dirac cones above the base energy, which is nothing but $\omega_e$ in our setup and is constrained by the chiral symmetry to be purely imaginary. We focus on the model studied in Ref.~\cite{Lee2019}: 
        \begin{multline}
        H_p(\mathbf{k})=J\sin (k_x) \sigma_x + J\sin (k_y) \sigma_y \\
        + iJ(2\cos (k_x) + \cos (k_y)-3)\,,
        \label{2DAIII}
        \end{multline}
        whose complex spectrum is shown in the left panel in Fig.~\ref{fig:NH2D}. With $\omega_e$ chosen to be $-iJ$, the topological number is $1$ since there is a single Dirac cone  with a positive chiral charge above $\omega_e$. The corresponding emitter Hamiltonian can then be obtained from Eq.~(\ref{AtomicBlochPiIdentity}). Its spectrum is shown in the right panel of Fig.~\ref{fig:NH2D}. The same reasoning gives the topological number of the emitter Hamiltonian as  $1$, implying the preservation of topology.

    \subsection*{Fewer-emitters-than-resonators case: violation of topological correspondence}

        We  consider now the most general case of a translationally invariant system of QEs coupled to a photonic lattice. 
		The interaction Hamiltonian within a unit cell can be written as 
		\begin{equation}
		    \hat H_{\mathrm{int},n}  = g\,\hat{\mathcal{S}}_n^\dag\,\Pi\,\hat{\mathcal{A}}_n + \mathrm{H.c.}
		\end{equation}
		where $\hat H_{\T{int}} = \sum_n \hat H_{\T{int},n}$, $\hat{\mathcal{A}}_n^{\T{\sc T}} = (\hat a_{n1},\ldots,\hat a_{n N_{s}})$, $\hat{\mathcal{ S}}_n^{\T{\sc T}} = (\hat \sigma_{n1},\ldots,\hat \sigma_{n N_{b}})$, c.f.~Eq.~\eqref{interaction}. 
		The rank of the projector $\Pi = \diag(p_1,\ldots,p_{N_b})$, $p_i\in\{0,1\}$, specifies the number of QEs  per unit cell. In the previous part we had $\Pi=\mathbb{1}_{N_b}$.
		
		The Bloch atomic Hamiltonian in this general case reads
		\begin{equation}\label{AtomicBlochPnotiIdentity}
		H_a(\textbf{k}) = 
		\Pi 
		\left(
		\omega_e
		+ \frac{g^2}{\omega_e-H_p(\textbf{k})}
		\right)
		\Pi\,,
		\end{equation}
		where only the non-zero block selected by $\Pi$ is relevant.
		Importantly, when the lattice constant of the emitter ``superlattice'' is larger than that of the photonic bath, the same expression for the effective Hamiltonian can be used, considering a suitably enlarged bath unit cell.

        By considering specific setups, we conclude that no general statements as in the one-emitter-per-resonator case can be made. In particular, we show that an Hermitian topological photonic lattice in 1D can induce both non-topological and topological interactions, according to the emitters' arrangement. Finally, we show how a  non-Hermitian non-topological photonic lattice can mediate non-Hermitian topology at the atomic level. These counterexamples support our claim.

        		\begin{figure*}[t!]
			\centering
			\includegraphics[width=17.5cm]{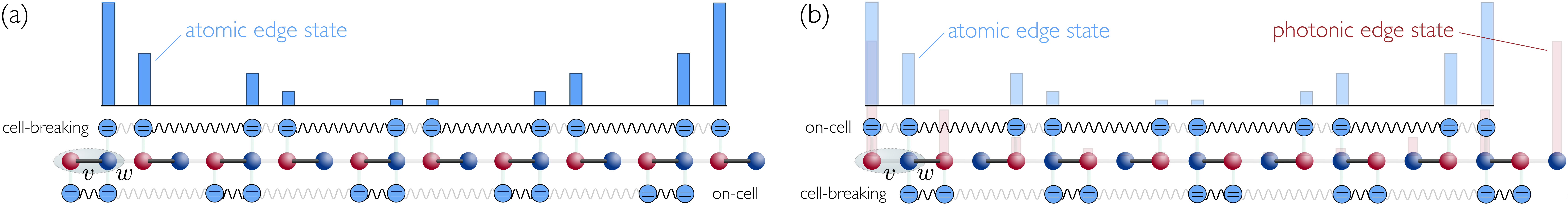}
			\caption{
				{\textit{Violation of the Hermitian topological preservation.}
					(a) Photonic lattice as in Eq.~\eqref{thetamodel} in real space under open BCs in the non-topological phase $v>w$. QEs are coupled to it in the on-cell and  cell-breaking configurations below and above the photonic lattice, respectively. 
					In the former, the atomic system inherits all symmetries and topology of the photonic lattice. 
					In the latter,
					a pair of topologically protected atomic edge states appear (only one is shown in cyan) even without photonic topology. For $\theta=0$, the atomic system has chiral symmetry and belongs to the BDI symmetry class, its topological number being the winding number. 
					Wiggly lines sketch the mediated couplings between QEs, their thickness representing the strength. In the almost-fully dimerized limit ($v\gg w$), the atomic system follows the photonic dimerization in the on-cell configuration, while it does not in the cell-breaking one, motivating the emergence of the edge states.
					(b) Same as (a), with a topological photonic lattice, with on-cell and cell-breaking configurations being above and below the photonic lattice, respectively. Here photonic edge states (shaded red) come along with atomic ones (shaded cyan) in the on-cell configuration. 
					Parameter values for the edge states: $\omega_e=0$, $\theta=0$ (a different $\theta$ does not substantially affect their profile), $w/v=4/5$ in (a), $v/w=2/5\, [4/5]$ in (b) for the photonic [atomic] one. 
				}
			} 
			\label{fig4}
		\end{figure*}

    \subsubsection*{Violation of Hermitian topological preservation}

    Consider a 1D photonic lattice whose Bloch Hamiltonian is
		\begin{multline}\label{thetamodel}
		H_p(k) = [v + w\cos k]\sigma_x \\
		            + w\sin k[\cos(2\theta)\sigma_y+\sin(2\theta)\sigma_z] \,,
		\end{multline}
		with $v,w>0$, $\theta\in[0,\pi/4]$. 
		This two-band model corresponds to an SSH lattice for $\theta=0$, and to a specific configuration of a Creutz ladder for $\theta=\pi/4$~\cite{LeonfortePRL2021}. 
		More precisely, $H_p(k)=U_\theta^\dagger H_\text{\tiny SSH}(k)U_\theta$, with $U_\theta=\cos\theta\mathbb{1}_2+i\sin\theta\sigma_x$ being a unitary transformation of the unit-cell modes. 
		It belongs to the BDI symmetry class for any $\theta$, $U_\text{\tiny TRS}=\cos(2\theta)\mathbb{1}_2-i\sin(2\theta)\sigma_x$, $U_\text{\tiny PHS}=\sigma_z$ and $U_\text{\tiny CH}=U_\text{\tiny PHS}U_\text{\tiny TRS}^* $   being the unitaries implementing antiunitary TRS and PHS, and unitary chiral symmetry, respectively~\cite{ChiuRMP2016,OzawaRMP2019}. 
		
		The original two-site photonic unit cell $(\hat a_{n,1},\hat a_{n,2})$ can be formally enlarged to a four-site one by relabeling $(\hat a_{n,1}, \hat a_{n,2}, \hat a_{n+1,1}, \hat a_{n+1,2}) \rightarrow (\hat a_{n,1}, \hat a_{n,2}, \hat a_{n,3}, \hat a_{n,4})$. 
		A two-band atomic Hamiltonian is then obtained by coupling only two QEs to two resonators in this enlarged photonic unit cell. 
		There are only two non-equivalent ways to do so.
		We set $\omega_e=0$, as to preserve chiral symmetry.
  
		i) The \emph{on-cell} case corresponds to coupling QEs to the resonators in sublattices 1 and 2, i.e., $\Pi = \diag(1, 1, 0, 0)$. 
	    Since $[\Pi, U_s] = 0$, for $s \in \{\T{TRS, PHS, CH}\}$, the atomic subsystem inherits all the symmetries of the photonic one. 
	    Besides belonging to the same symmetry class, atomic and photonic systems have the same topology, as in the one-emitter-per-resonator case.
	    Indeed, by highlighting the dependence on the parameters in the Bloch photonic Hamiltonian $H_\text{\tiny SSH}(k; v, w)$, one can see that $H_a(k)\propto [H_\text{\tiny SSH}(k; v^2, -w^2)]^{-1}$.
	    Thus, the atomic and photonic systems possess the same winding number~\cite{AsbothBook2016}.
	    Note that it is sufficient to analyze the $\hat H_a$ obtained in the SSH case ($\theta = 0$) since the other cases ($\theta \neq 0$) are unitarily equivalent.
		    
		ii) The \emph{cell-breaking} case corresponds to coupling QEs to the resonatorsin sublattices 2 and 3, i.e., $\Pi = \diag(0, 1, 1, 0)$. The atomic Hamiltonian inherits PHS, with $U_\text{\tiny PHS} = -\sigma_z$ for any $\theta$, but loses TRS and chiral symmetry, except for $\theta=0$. 
	    For an SSH photonic lattice ($\theta=0$), the mediated Hamiltonian is topological when the photonic one is not, and vice versa.
	    Indeed, $H_a(k) \propto [H_\text{\tiny SSH}(k; w^2, -v^2)]^{-1}$, therefore the photonic winding number is zero (non-zero) when the atomic one is non-zero (zero).
	    This is reminiscent of the change in topology following a redefinition of the intra/intercell coupling amplitudes.
	    For $0 < \theta < \pi/4$, the mediated Hamiltonian belongs to the D symmetry class, having only PHS, while for $\theta=\pi/4$ the mediated atomic Hamiltonian is gapless for any $v, w$, the gap being $\Delta = 2 g^2 w\,\cos(2\theta)/(v^2 + w^2)$. It is therefore meaningless to compare the photonic and atomic topologies in these cases.

    \subsubsection*{Violation of non-Hermitian topological reversal}

     Just like the Hermitian case, we cannot generally claim topological reversal or preservation if the number of emitters is smaller than that of resonators. 
     A general method to construct counterexamples is that of stacking two topological photonic lattices with opposite topological numbers and coupling only one of them to emitters in the one-emitter-per-resonator manner.  
     While the entire photonic bath is trivial, the emitter Hamiltonian will be nontrivial due to either topological reversal or preservation with respect to the sublattice with which it interacts.
     The nontrivial emitter topology  persists even if we turn on the coupling between the two photonic lattices, provided the coupling is not so strong that the emitter Hamiltonian remains gapped. Note that this recipe also applies to Hermitian systems. 
        
\begin{figure}[t!]
			\centering
			\includegraphics[width=0.95\columnwidth]{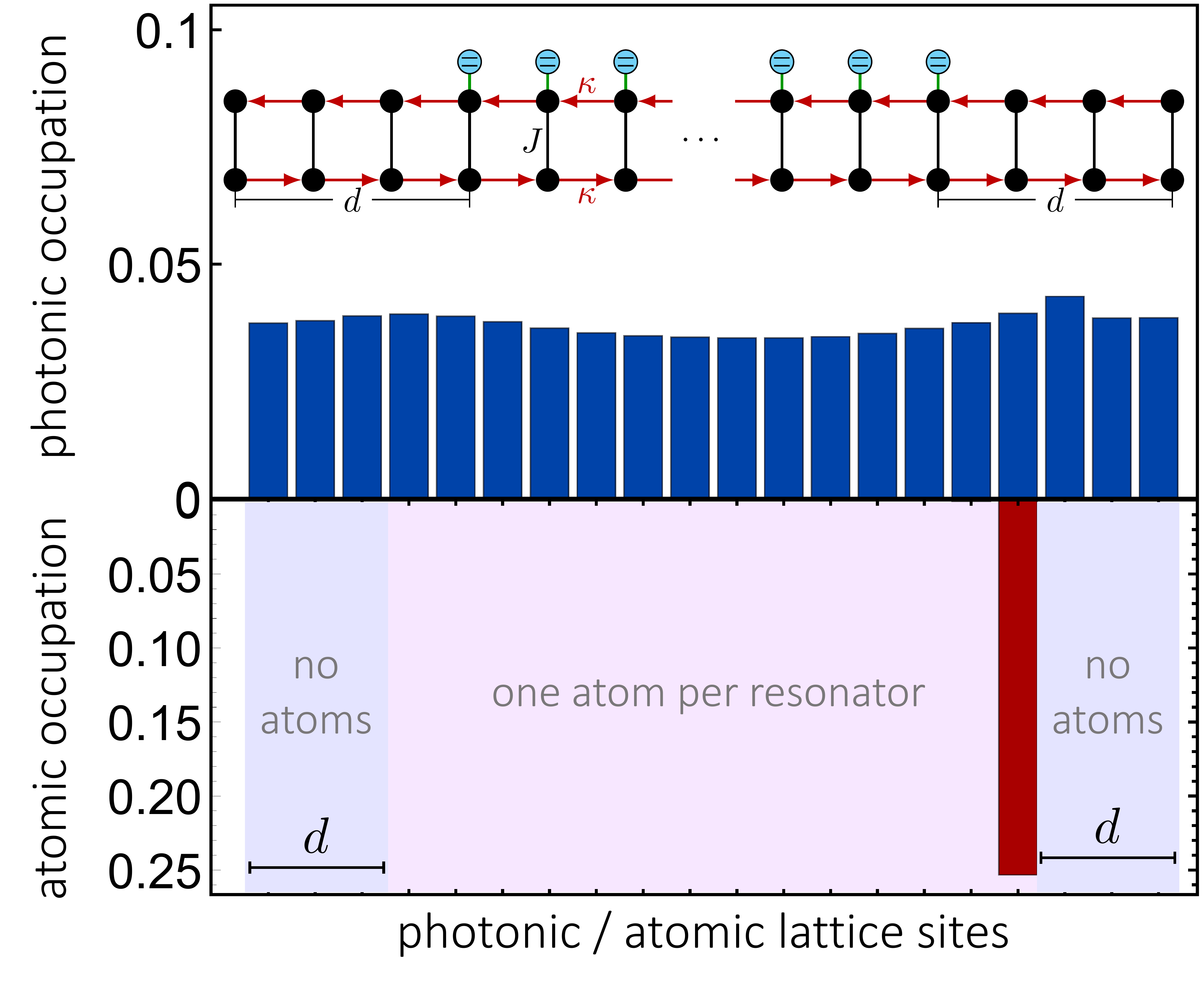}
			\caption{
				{\textit{Violation of the non-Hermitian topological reversal.} The photonic bath is built from two coupled unidirectional Hatano-Nelson chains with opposite chiralities with $2N$ sites in total. Under  open BCs, only the resonators in the upper chain are coupled to emitters, except for the leftmost and rightmost $d$ sites. Note that a background loss captured by an imaginary shift $-i\kappa$ is omitted in the inset. There is no photonic skin effect (top), yet the atomic skin effect occurs (bottom). Both figures display $\overline{|\langle n|\psi_i\rangle|^2}$ with $n$ being an atomic or photonic site and $\overline{\cdots}$ being the average over all the normalized right eigenstates $|\psi_i\rangle$. On the photonic side, we further sum up each pairs of sites in the same unit cell, i.e., those with the same horizontal positions. Here the parameters are chosen to be $N=20$, $d=3$, $J=0.5\kappa$ and $g=0.1\kappa$.
                }
			}
			\label{fig:CE} 
		\end{figure}

     As a simple (counter)example, we construct a 1D non-Hermitian photonic bath with zero spectral winding number that nevertheless induces a nonzero spectral winding number for the emitter Hamiltonian. The bath consists of two unidirectional Hatano-Nelson chains with opposite chiralities and Hermitian inter-chain hopping. The emitters are only coupled to one of them, say the leftward one. We can then write down the photonic Bloch Hamiltonian and projector as
        \begin{equation}
        H_p(k)= 
        \begin{bmatrix}
        \kappa(e^{ik}-i) & J \\
        J & \kappa(e^{-ik}-i)
        \end{bmatrix},\;\;\;\;
        \Pi=\begin{bmatrix}
        1 & 0 \\
        0 & 0
        \end{bmatrix}.
        \label{HpPi}
        \end{equation}
        We assume $J<\kappa$ so that $H_p(k)$ is gapped with respect to $\omega_e=-i\kappa$. Substituting the above expressions (\ref{HpPi}) into Eq.~(\ref{AtomicBlochPnotiIdentity}), we obtain
        \begin{equation}
        H_a(k)=-\frac{g^2\kappa}{\kappa^2 - J^2}e^{-ik} - i\kappa,
        \label{uniHa}
        \end{equation}
        which turns out to be characterized by a nonzero spectral winding number $\nu_a=-1$ with respect to the base energy $\omega_e$~\cite{Gong2018}. To visualize such a breakdown of non-Hermitian topological reversal in this model, we consider a finite system under  the open boundary condition and cut off some emitters near the boundaries. As shown in Fig.~\ref{fig:CE}, we find that the skin effect occurs in the emitter array, despite the fact that the photonic bath does not exhibit a skin effect.

    \section*{Discussion}

        We have investigated  the relationship between the topological properties of a photonic lattice and those of the photon-mediated dipole-dipole Hamiltonian of a periodic set of atoms coupled to it.

        A complete characterization is carried out in the case of a topological photonic lattice, both in the Hermitian and non-Hermitian case, where each mode is coupled to a single quantum emitter. The atomic system, under a resonance condition maintaining the photonic symmetries, can  possess either the same or opposite topology depending on the spatial dimension and  whether the photonic Hamiltonian is Hermitian or not. 

        Within the general classification we provide, the topological reversal stands out in particular. In Hermitian systems, this effect predicts the existence of atomic boundary modes featuring excitations propagating with opposite chirality with respect to the photonic ones of the underlying 2D lattice. In 1D non-Hermitian systems instead, our analysis provides a general framework for understanding reversed non-reciprocal mediated interactions by non-reciprocal photonic baths~\cite{RoccatiOptica2022,GongPRL2022}.

        The case of a fully translationally invariant setup, though with fewer emitters than resonators, has been also considered. A reduced number of atomic degrees of freedom represent a weaker constraint so that no general photon-to-atom topological correspondence is found. To support our claim we have discussed two instances of the violation of such correspondence.

    \section*{Methods}

		\subsection*{Photonic Hamiltonian
		}
		
		The photonic Hamiltonian in real space reads
		\begin{equation}
		    \hat H _p
		    =
		    \sum_{n,m=1}^{N_c}
		    \sum_{s,s'=1}^{N_b}
		    \bra{\textbf{r}_n,s}
		    \hat H _p
		    \ket{\textbf{r}_{n+m},s'}
		    \hat a_{ns}^\dagger
		    \hat a_{n+m,s'},
		\end{equation}
		where $N_c$ is the number of unit cells, $\ket{\textbf{r}_{n},s}=\hat a_{ns}^\dagger \ket{\T{vac}}$ and the closure relation is $\mathbb{1}_p=\sum_{n,s} \dyad{\textbf{r}_{n},s}$. 
        Translational invariance impose the couplings to be independent on the cell position, 
		$\bra{\textbf{r}_n,s}
		    \hat H _p
		    \ket{\textbf{r}_{n+m},s'}=\bra{\textbf{r}_0,s}
		    \hat H _p
		    \ket{\textbf{r}_{m},s'}$.
		    
		Assuming 
        periodic BCs and using the closure relation
		$\mathbb{1}_p=\sum_{\textbf{k},s} \dyad{\textbf{k},s}$, where 
		$\ket{\textbf{k},s}=\hat a_{\textbf{k}s}^\dagger \ket{\T{vac}}$
		and the overlap $\braket{\textbf{r}_n,s}{\textbf{k},s'}=\delta_{ss'}e^{-i \textbf{k}\cdot\textbf{r}_n}/\sqrt{N_c}$, we have
		\begin{equation}
		    \hat a_{ns}
		    =
		    \frac{1}{\sqrt{N_c}}
		    \sum_{\textbf{k}}
		    e^{-i \textbf{k}\cdot\textbf{r}_n}
		    \hat a_{\textbf{k}s}\,.
		\end{equation}

		This gives the photonic Hamiltonian of the periodic lattice as
		\begin{equation}
		    \hat H _p
		    =
		    \sum_{\textbf{k}}
		    \sum_{s,s'=1}^{N_b}
		    %\sum_{m}
		    \bra{\textbf{r}_0,s}
		    \hat H _p
		    \ket{\textbf{r}_{n},s'}
		    e^{-i \textbf{k}\cdot\textbf{r}_n}
		    \hat a_{\textbf{k}s}^\dagger
		    \hat a_{\textbf{k}s'}\,.
		\end{equation}
		By introducing the vector operator 
		$\hat A_\textbf{k}^\dagger=(\hat a_{\textbf{k},1}^\dagger,...,\hat a_{\textbf{k},N_b}^\dagger)$
		and denoting the matrix elements
		\begin{equation}
            [H_p(\textbf{k})]_{ss'}
            =
            \sum_n
            \bra{\textbf{r}_0,s}
		    \hat H _p
		    \ket{\textbf{r}_{n},s'}
            e^{-i\textbf{k}\cdot \textbf{r}_n}\,,
        \end{equation}
		it can be recast in the form given in the main text, c.f.~Eq.~\eqref{HB}.

		\subsection*{Effective mediated Hamiltonian}
		Consider the atomic frequency $\omega_e$ to be at distance 
		$\Delta$ from the photonic continuum. If the atom-photon coupling $g$ is weak so that $g/\Delta\ll1$, then it is possible to adiabatically eliminate the 
		photonic bath and derive an effective photon-mediated atomic Hamiltonian $\hat H_a$~\cite{CohenAP,SanchezPRA2020}.
		
		In the case of one emitter per resonator, its explicit expression reads
		\begin{equation}\label{Hamethods}
		    \hat H_a
		    =
		    \hat H_e
		    +
		    \sum_{nm}\sum_{ss'} h_{ns,ms'}\hat \sigma^\dagger_{ns}\hat \sigma_{ms'}
		\end{equation}
		where 
		\begin{equation}\label{hnm}
		    h_{ns,ms'}
		    =
		    g^2\bra{\textbf{r}_m,s'}\hat G_p(\omega_e)\ket{\textbf{r}_n,s}
		\end{equation}
		with $\hat G_p(z)=(z-\hat H_p)^{-1}$ being the Green's function of the bare photonic Hamiltonian and 
		$\ket{\textbf{r}_m,s}$
		the state with one excitation in the $s$th resonator of the $m$th unit cell of the photonic lattice. The double index in the atomic operators specifies both cell ($n,m$) and sublattice ($s,s'$) the emitter is coupled to.
		
		As $\hat H_p$ is translationally invariant, so is its resolvent operator~\cite{Economou2006} and
		\begin{equation}\label{greentrasl}
		    \bra{\textbf{r}_m}\hat G_p(\omega_e)\ket{\textbf{r}_n}
		    =
		    \frac{1}{N}
		    \sum_{\textbf{k}\in \T{BZ}}
		    \frac{e^{i\textbf{k}\cdot (\textbf{r}_m-\textbf{r}_n) }}{\omega_e - H_p(\textbf{k})}
		\end{equation}
		where 
		$\bra{\textbf{r}_m}\hat G_p(\omega_e)\ket{\textbf{r}_n}$
		is the $N_b\times N_b$ matrix in the sublattice space.
		Therefore, the atomic Hamiltonian can be written in Bloch form as 
		\begin{equation}\label{HaB}
		\hat H_a= \sum_{\textbf{k}\in \T{BZ}}\hat S_\textbf{k}^\dagger\, H_a(\textbf{k})\hat S_\textbf{k}
		\end{equation}
		where  $\hat S_\textbf{k}^\dagger=(\hat \sigma_{\textbf{k}1}^\dagger,...,\hat \sigma_{\textbf{k}N_b}^\dagger)$, $H_a(\textbf{k})$ is the Bloch  Hamiltonian as in Eq.~\eqref{AtomicBlochPiIdentity} and 
		\begin{equation}
		    \hat \sigma _{\textbf{k}s} 
		    =
		    \frac{1}{\sqrt{N}}
		    \sum_n
		    e^{i\textbf{k}\cdot \textbf{r}_n}
      \hat \sigma_{ns}
		\end{equation}
        with
        $\textbf{r}_n$ being the atomic operator position in real space~\cite{GongPRA2022}.
		
		In the case of fewer emitter than resonators, still arranged so to preserve translational invariance, one can repeat the above arguments, with a small modification: the indices $s$ and $s'$ in Eq.~\eqref{hnm} belong only to the sublattices coupled to QEs, i.e. those for which $p_1=1$ in the projector $\Pi=\T{diag}(p_1,\ldots,p_{N_b})$. This straightforwardly leads to the insertion of the projection $\Pi$ in Eq.~\eqref{greentrasl}, which yields Eq.~\eqref{AtomicBlochPnotiIdentity} in the main text.

        \subsection*{Triviality of the full atom-light Hamiltonian}
        
        Here we prove that the entire atom-light Hamiltonian in Eq.~\eqref{Hes} is topologically trivial.
        Its spectrum and eigenstates can be computed analytically as follows. 
        Suppose $U_\mathbf{k}$ is the unitary that diagonalizes $H_p(\mathbf k)$, $U^\dag_\mathbf{k} H_p(\mathbf k) U_\mathbf{k} = \diag\left(\omega_1(\mathbf k),\omega_2(\mathbf k),\dots,\omega_N(\mathbf k)\right)\equiv \Lambda(\mathbf k)$. 
        Then, 
        \begin{equation}
            (I_2\otimes U_\mathbf{k})^\dag H(\mathbf k) (I_2\otimes U_\mathbf{k}) = \begin{bmatrix}
            \omega_e I & g I \\ g I & \Lambda(\mathbf k)
            \end{bmatrix} \,.
        \end{equation}
        Thus, for each band of the bare bath, $\omega_j(\mathbf k)$, there are two bands
        \begin{equation}
            \omega_{\pm,j} = \frac{\omega_e + \omega_j(\mathbf k)}{2} \pm \sqrt{\frac{[\omega_e - \omega_j(\mathbf k)]^2}{4} + g^2} \,,
        \end{equation}
        which are eigenvalues of 
        \begin{equation}
            H_j(\mathbf k) = \frac{\omega_e + \omega_j(\mathbf k)}{2} I + \frac{\omega_e - \omega_j(\mathbf k)}{2} \sigma_z + g \sigma_x \,.
        \end{equation}
        And the corresponding eigenvectors are $\ket{v_{\pm, j}(\mathbf k)}\otimes\ket{u_j(\mathbf k)}$, where $\ket{u_j(\mathbf k)}$ is the eigenstate of $H_p(\mathbf k)$ with eigenvalue $\omega_j(\mathbf k)$, while $\ket{v_{\pm,j}(\mathbf k)}$ is the eigenstate of $H_j(\mathbf k)$ with eigenvalue $\omega_{\pm, j}(\mathbf k)$.
        Note that, regardless the value of $\omega_e$, as long as it lies in a gap of the bare bath's spectrum, half of the spectrum is above it and half below it, $\omega_{-,j} < \omega_e < \omega_{+,j}$ for all $j$.
        If we now consider the bands below $\omega_e$ and compute the topological invariant, we can consider instead the topologically equivalent Hamiltonian
        \begin{equation}
            H = I - 2 P(\mathbf k) \,,
        \end{equation}
        with $P(\mathbf k) = \sum_j \ket{u_j(\mathbf k)}\!\bra{u_j(\mathbf k)}\otimes \ket{-}\!\bra{-} = I \otimes \ket{-}\!\bra{-}$, with a constant $\ket{-}$, therefore $dH = 0$, so ${\rm Ch}_n = 0$, c.f.~Eq.~\eqref{Chnumber}.
        
        For chiral systems in odd dimensions, the bath's Bloch Hamiltonian can be written as 
        \begin{equation}
            H_p(\mathbf k) = \begin{bmatrix}
            0 & Q_p(\mathbf k) \\ Q_p^\dag(\mathbf k) & 0
            \end{bmatrix} \,,
        \end{equation}
        with $Q_p(\mathbf k)$ being a suitable matrix.
        Then, the Bloch Hamiltonian of the bath with emitters ($\omega_e = 0$) can also be written in the same block-off-diagonal form, with
        \begin{equation}
           Q(\mathbf k) = \begin{bmatrix}
           Q_p(\mathbf k) & g I \\ g I & 0
           \end{bmatrix} \,.
        \end{equation}
        Note that the inverse is given by
        \begin{equation}
            Q(\mathbf k)^{-1} = \begin{bmatrix}
            0 & g^{-1} I \\ g^{-1}I & -g^{-2} Q_p(\mathbf k)
            \end{bmatrix} \,.
        \end{equation}
        Thus,
        \begin{equation}
            Q^{-1} d Q = \sum_j \begin{bmatrix}
            0 & 0 \\ g^{-1}\partial_j Q & 0
            \end{bmatrix} dk_j \,.
        \end{equation}
        As a consequence, $\Tr[(Q^{-1} dQ)^{2n + 1}] = 0$, so $\nu_{2n + 1} = 0$, c.f.~Eq.~\eqref{wnumber}.

        Finally, we provide an alternative proof showing that the entire system is trivial without referring to any formulas of topological invariants. We note that Eq.~(\ref{Hes}) can be continuously deformed into 
        \begin{equation}
        H_1=(\omega_e\sigma_0+g\sigma_x)\otimes I\,,
        \end{equation}
        with $\sigma_0$ being the two-by-two identity matrix,
        via a linear interpolation 
        \begin{equation}
        H_\lambda(\mathbf k)=(1-\lambda)H(\mathbf k)+\lambda H_1,\;\;\;\; \lambda\in[0,1]. 
        \end{equation}
        One can check that $\det(H_\lambda(\mathbf k)-\omega_e\sigma_0\otimes I)=\det(-g^2I)\neq0$, so the Hamiltonian remains gapped near $\omega_e$ during the deformation. Note that any time-reversal symmetry is preserved, so is the particle-hole (chiral) symmetry if it is extended as $(-C)\oplus C$ ($(-S)\oplus S$). Since $H_1$ does not depend on $\mathbf k$ and is thus obviously trivial, we conclude that $H(\mathbf k)$, which is continuously connected to $H_1$, is also trivial. Note that the above proof applies equally to Hermitian and non-Hermitian systems. Moreover, the fact that an appropriately extended chiral symmetry requires a minus sign on the emitter side explains why the triviality of the entire system does not contradict topological preservation in chiral symmetric systems.

        \subsection*{Proof of topological preservation and reversal}
        Here we provide a general analysis of which fundamental symmetry class (in Hermitian AZ, non-Hermitian AZ, or non-Hermitian AZ$^\dag$) exhibits topological reversal or otherwise topological preservation for the one-emitter-per-resonator setup. As in the main text, to ensure that the PHS and chiral symmetry are inherited, we focus on the case of $\omega_e=0$, so that $H_a(\textbf{k})=-g^2H_p(\textbf{k})^{-1}$ with both $H_a(\textbf{k})$ and $H_p(\textbf{k})$ gapped near $0$. 
        We observe that for the non-Hermitian case, both Bloch Hamiltonians have to be point-gapped around $\omega_e$ with a negative imaginary part so that their spectra both lie below the real axis in the complex energy plane. This is just a rigid shift along the imaginary axis that does not affect the following discussion. 

        One obvious observation is that the mapping from $H_p(\textbf{k})$ to $H_a(\textbf{k})$ is invertible. This immediately implies that, after taking the topological equivalence classes of $H_a(\textbf{k})$ and $H_p(\textbf{k})$, we obtain an automorphism on the classification group. Recalling that nontrivial Hermitian AZ classes are classified by $\mathbb{Z}_2$ or $\mathbb{Z}$,\footnote{In the literature, the topological classifications of some classes are usually denoted as $2\mathbb{Z}$, meaning that the winding number or Chern number can only be an even integer. Nevertheless, since $2\mathbb{Z}$ is isomorphic to $\mathbb{Z}$, the convention $\mathbb{Z}$ is also used.} and so are the non-Hermitian AZ (AZ$^\dag$) classes, it suffices to consider the automorphisms on $\mathbb{Z}_2$ or $\mathbb{Z}$ (with respect to addition). In the former case ($\mathbb{Z}_2$), the only automorphism is the identity map, implying that all the $\mathbb{Z}_2$ phases exhibit topological preservation. In the latter case ($\mathbb{Z}$), the only two possibilities of an automorphism are identity map and inversion ($n\mapsto-n$), corresponding to topological preservation and switch, respectively. We emphasize the above results apply to both Hermitian and non-Hermitian systems. The remaining problem is thus to distinguish the $\mathbb{Z}$ phases exhibiting topological reversal from those exhibiting topological preservation. 
        
        We first focus on the Hermitian case. Using the band-flattening ($H\to {\rm sgn}H$) technique \cite{Kitaev2009}, we know that the map from $H_p(\textbf{k})$ to $H_a(\textbf{k})$ can be simplified into a simple inversion ($H\to -H$) on the level of topological equivalence classes. If the spatial dimension is odd, all the $\mathbb{Z}$ phases are chiral symmetric, and thus the Bloch Hamiltonian takes the form 
        \begin{equation}
        H(\textbf{k})= 
        \begin{bmatrix} 
        0 & Q(\textbf{k}) \\
        Q(\textbf{k})^\dag & 0 
        \end{bmatrix}.
        \end{equation}
        The integer topological number is the winding number given by
        \begin{equation}
        \nu\propto \int_{\rm BZ}\Tr(Q^{-1}dQ)^D\,.
        \label{wnumber}
        \end{equation}
        Obviously, this topological invariant does not change upon the inversion $H\to-H$ (leading to $Q\rightarrow-Q$). Otherwise, in even spatial dimensions, the topological number is the Chern number, which is determined by the flattened Bloch Hamiltonian $H(\textbf{k})$ via 
        \begin{equation}
        {\rm Ch}\propto \int_{\rm BZ}\Tr(H(dH)^D).
        \label{Chnumber}
        \end{equation}
        Unlike the winding number, the Chern number is also inversed upon the inversion of the Hamiltonian. 

        We move now  to the non-Hermitian case. Here the counterpart of band flattening is unitarization~\cite{Gong2018}
        \begin{equation}
            H\to  V=H\left(\sqrt{H^\dag H}\right)^{-1}\,,
        \end{equation} 
        upon which the photon-to-atom map is simplified into $V\to -V^\dag$. As mentioned in the main text, the topological numbers in odd dimensions are the winding numbers given in Eq.~\eqref{wnumber} and we always have the topological reversal. In even dimensions, any $\mathbb{Z}$ topological phase exhibits a chiral symmetry $S$ (cf. Eq.~\eqref{NHCS}). The integer topological number is then given by the Chern number \eqref{Chnumber} of $iSV$~\cite{KawabataPRX2019}, which can be checked to be Hermitian and flattened (i.e., square to identity). After the operation $V\to-V^\dag$, this quantity turns out to undergo a unitary conjugation: 
        \begin{equation}
           iSV\to -iSV^\dag = iVS= S^{-1}(iSV)S,
        \end{equation}
        leaving the Chern number unchanged.

        In summary, concerning the Hermitian AZ classes, a topological reversal occurs only for $\mathbb{Z}$ phases in even dimensions. Concerning the non-Hermitian AZ (AZ$^\dag$) classes, a topological reversal occurs only for $\mathbb{Z}$ phases in odd dimensions.

        \vspace{5mm}

        \textit{Acknowledgments---} This work was partially funded by the Luxembourg National Research Fund (FNR, Attract grant 15382998). M.~B.~and Z.~G.~acknowledge financial support from the MCQST. Z.~G.~was supported by the Max-Planck-Harvard Research Center for Quantum Optics (MPHQ). F.~C.~acknowledges support from Ministero dell’Universit\`a e della Ricerca (MUR) through project PRIN (Project No. 2017SRN-BRK QUSHIP). We acknowledge Diego Porras and  Pablo Mart\'inez-Azcona for useful discussions.

		\bibliographystyle{apsrev4-1}			
		\bibliography{NHtopologyWQED}
  
  \end{document}